\documentclass[reprint,superscriptaddress,amsmath,amssymb,aps,prl,longbibliography]{revtex4-2}

\usepackage[utf8]{inputenc}
\usepackage[english]{babel}
\babeladjust{bcp47.toname=on}
\DeclareUnicodeCharacter{03B2}{\ensuremath{\beta}}

\usepackage{graphicx}
\usepackage{dcolumn}
\usepackage{bm}
\usepackage[colorlinks=true,urlcolor=blue,linkcolor=blue,citecolor=blue]{hyperref}
\usepackage[normalem]{ulem}
\usepackage{tikz}
\usetikzlibrary{quantikz}
\usepackage{amsmath}
\usepackage{amssymb}
\usepackage{bbold}
\usepackage{algorithm}
\usepackage{comment}
\DeclareMathOperator{\Tr}{Tr}
\usepackage{multirow}

\usepackage[all]{hypcap} 
\usepackage{braket}

\newcommand{\customref}[2]{\hyperref[#1]{\ref*{#1}#2}}

\usepackage{xcolor}

\definecolor{Ured}{HTML}{FF5C5C}
\definecolor{Ublue}{HTML}{ADD8E6}
\definecolor{Ugreen}{HTML}{198a11}

\renewcommand{\vec}[1]{\boldsymbol{#1}}

\begin{document}

\title{Syndrome resampling enhances quantum error correction thresholds}

\author{Luis Colmenarez}
\email{colmenarez@physik.rwth-aachen.de}
\affiliation{Institute for Theoretical Nanoelectronics (PGI-2), Forschungszentrum Jülich, 52428 Jülich, Germany}
\affiliation{Institute for Quantum Information, RWTH Aachen University, 52056 Aachen, Germany}

\author{Áron Márton}
\affiliation{Institute for Theoretical Nanoelectronics (PGI-2), Forschungszentrum Jülich, 52428 Jülich, Germany}
\affiliation{Institute for Quantum Information, RWTH Aachen University, 52056 Aachen, Germany}

\author{Markus M\"{u}ller}%
\affiliation{Institute for Theoretical Nanoelectronics (PGI-2), Forschungszentrum Jülich, 52428 Jülich, Germany}
\affiliation{Institute for Quantum Information, RWTH Aachen University, 52056 Aachen, Germany}

\date{\today}

\begin{abstract}
Quantum error correction (QEC) enables fault-tolerant quantum computation but requires operating quantum hardware at physical error rates below code-dependent thresholds, which remains challenging for current devices. We introduce \emph{syndrome resampling}, a general method that increases QEC thresholds of any decoder and suppresses logical errors without additional hardware, decoding modifications, or code-specific assumptions beyond syndrome statistics. The method exploits the fact that syndromes with low probability are likely to lead to logical failure, therefore biasing syndrome averages towards most likely syndromes effectively increases logical fidelities.  We establish a direct connection between the R\'enyi coherent information (RCI) and powers of the syndrome probability distribution, showing that resampling syndromes according to these powers combined with maximum likelihood decoding (MLD) realizes a family of optimal thresholds associated with phase transitions in the RCI. Numerical simulations of surface codes demonstrate that syndrome resampling substantially increases thresholds for both optimal and suboptimal decoders and reduces logical error rates by up to four orders of magnitude in experimentally relevant regimes. We further show that syndrome resampling can be effectively implemented from finite data and combined with decoding-based post-selection to achieve additional gains. Finally, applying the method to existing experimental QEC data yields up to two orders of magnitude reduction in logical error rates without requiring additional measurements. Our results provide a practical and decoder-agnostic route to improved logical fidelities in near-term QEC experiments. 
\end{abstract}

\maketitle

Quantum error correction (QEC) is a central ingredient for fault-tolerant (FT) quantum computation. By introducing redundant degrees of freedom, QEC enables the detection and correction of errors by encoding one or more logical qubits into many noisy physical qubits. Measurements revealing the presence of errors are referred to as \emph{syndrome}. Recent experiments in superconducting qubits \cite{google_quantum_ai_and_collaborators_quantum_2025,google_quantum_ai_and_collaborators_scaling_2025,google_quantum_ai_and_collaborators_magic_2025,krinner_realizing_2022,besedin_lattice_2026,caune_demonstrating_2026,gupta_encoding_2024,hetenyi_creating_2024,zhao_realization_2022,putterman_hardware-efficient_2025}, trapped ions \cite{yamamoto_quantum_2026,ryan-anderson_high-fidelity_2024,paetznick_demonstration_2024,butt_demonstration_2026,hong_entangling_2024,huang_comparing_2024,mayer_benchmarking_2024,pogorelov_experimental_2025,postler_demonstration_2022,postler_demonstration_2024,stricker_experimental_2020}, and neutral atoms \cite{bluvstein_logical_2024,reichardt_fault-tolerant_2025,sales_rodriguez_experimental_2025,bluvstein_fault-tolerant_2026,baranes_leveraging_2026} have demonstrated key QEC protocols, marking the onset of the early FT era.
Scaling QEC to large FT algorithms requires device noise to lie below certain \emph{thresholds} \cite{shor_scheme_1995,shor_fault-tolerant_1996,knill_resilient_1998}. While several experiments report operation below these thresholds \cite{google_quantum_ai_and_collaborators_quantum_2025,sivak_real-time_2023,paetznick_demonstration_2024,gupta_encoding_2024,putterman_hardware-efficient_2025,bluvstein_fault-tolerant_2026}, further progress demands both more physical qubits and lower noise levels. It is therefore desirable to suppress logical errors using methods that do not require substantial additional hardware resources.
One route to improving logical fidelities is improved decoding \cite{terhal_quantum_2015,iolius_decoding_2024}. However, decoding has fundamental limits: the Maximum Likelihood Decoder (MLD) \cite{terhal_quantum_2015,fuentes_degeneracy_2021} achieves optimal performance, but solving MLD is computationally intractable in general \cite{iyer_hardness_2015}. These limits are reflected in a phase transition of the coherent information of the noisy quantum state associated with the QEC code \cite{schumacher_quantum_1996,fan_diagnostics_2024,colmenarez_accurate_2024,huang_coherent_2025,niwa_coherent_2025,colmenarez_fundamental_2025,dennis_topological_2002,chubb_statistical_2021,vodola_fundamental_2022,rispler_random_2026,hauser_information_2026}.
An alternative approach is post-selection (PS), in which any runs exhibiting detected errors are discarded, thereby avoiding decoding altogether. However, the probability of obtaining error-free runs decays exponentially with the number of faulty operations \cite{aliferis_accuracy_2008}. Recently, intermediate strategies between decoding and PS have been proposed, including confidence-based decoding \cite{smith_mitigating_2024,dinca_error_2026,lee_efficient_2026,bombin_fault-tolerant_2024,xie_simple_2026,meister_efficient_2024}, heuristic post-selection rules \cite{english_thresholds_2025,chen_scalable_2025,staples_scalable_2026,birchall_macromux_2026}, and different forms of logical error mitigation \cite{aharonov_syndrome_2025,jeon_quantum_2026,suzuki_quantum_2022,piveteau_error_2021,lostaglio_error_2021,xiong_sampling_2020,zhang_demonstrating_2025,zhou_error_2025,umbrarescu_infinite_2026}. These methods are typically tailored to specific codes and decoders, require additional overhead, and lack a unifying theoretical framework to assess their impact on QEC thresholds.

In this work, we establish a connection between the Rényi Coherent Information (RCI) and powers of the syndrome probability distribution (SPD). Based on this relation, we introduce \emph{syndrome resampling} (SR), a general method to enhance thresholds and substantially improve logical fidelities, specially in the regime of error rates of current devices. We identify a family of thresholds associated with phase transitions in the RCI, which characterize the optimal performance attainable after SR. We show that constructing the SPD and applying SR can significantly increase the threshold for both optimal and sub-optimal decoders, while reducing logical error rates by up to four orders of magnitude.
We call the enhanced thresholds under SR \emph{Rényi thresholds} to distinguish them from standard fault-tolerant thresholds~\cite{iyer_hardness_2015}.
Unlike previous methods used to increase QEC thresholds \cite{smith_mitigating_2024,english_thresholds_2025}, our method does not rely on decoding confidence or code-specific information beyond the syndrome statistics. 
We benchmark SR in a setting where the SPD can be computed exactly and outline a practical scheme to estimate it from syndrome measurements. While the number of required samples can be large, it depends on the noise strength and the desired accuracy of the SPD estimate. We further demonstrate that combining SR with PS based on the complementary gap~\cite{smith_mitigating_2024} yields additional gains in logical fidelity in experimentally relevant regimes.
Finally, we apply SR to experimental QEC data from a recent demonstration of lattice surgery \cite{besedin_lattice_2026} and observe a reduction of the logical error rate by up to two orders of magnitude, without requiring additional syndrome measurements, modifications to the decoding algorithm and keeping one order of magnitude more samples than PS.

\emph{The Rényi Coherent Information (RCI) and powers of the syndrome probability distribution (SPD)}:
A QEC code with parameters $[[n,k,d]]$ is defined as the common eigenspace of the stabilizer generators $S_i$, with $i=1,..,n-k$. 
We choose the codespace corresponding to the $+1$ eigenspace of all stabilizers.
The logical operators $(O^X_{L_j},O^Z_{L_j})$, with $j=1,..,k$, represent the Pauli $X$ and $Z$ acting on the $k$ logical qubits encoded in the QEC code. 
We consider a maximally entangled state $\rho^0_{RQ}$ between the $k$ logical qubits, supported on the $n$-qubit physical system $Q$, and a $k$-qubit reference system $R$ \cite{fan_diagnostics_2024,colmenarez_accurate_2024,colmenarez_fundamental_2025}. As such, this state is a stabilizer state given by the common state of the $S_i=+1$ stabilizers and all products $O^{X,Z}_{R_j} O^{X,Z}_{L_j} = +1$. Equivalently, it can be written in the stabilizer basis  as $\rho^0_{RQ}= |s_0,B_0\rangle \langle s_0,B_0 | $, where $s_0$ denotes the trivial syndrome and $B_0$ labels a Bell basis state between reference and logical qubits. 
Now we define the RCI as 
\begin{equation}\label{eq:def_coherent_information}
    I^{(\alpha)} = S^{(\alpha)}(\rho_Q) - S^{(\alpha)}(\rho_{RQ}).
\end{equation}
Here, $S^{(\alpha)}(\rho)=\log (\Tr\rho^{\alpha})/(1-\alpha)$ is the $\alpha-$th Renyi entropy of the state and $\rho_Q = \Tr_R{(\rho_{RQ})}$ is the reduced density matrix after tracing out the reference system. In general $\alpha\geq 0$, with the limit $\alpha\rightarrow 1$ recovering the von Neumann entropy and therefore the standard CI \cite{schumacher_quantum_1996,colmenarez_accurate_2024}.
The noisy map $\mathcal{N}(\rho) = \sum_P P(E)E \rho E^{\dagger}$ is assumed to be Pauli noise such that $E \in \mathcal{P}$ is an element of the $n$-qubit Pauli group and $P(E)$ is the probability of each Pauli error operator $E$. Since $\rho_{RQ}^0$ is a pure stabilizer state then $\rho_{RQ}$ is a mixed stabilizer state diagonal in the stabilizer basis \cite{colmenarez_accurate_2024,bausch_error_2021} of the respective error correcting code:
\begin{eqnarray}
    \rho_{RQ}  = \sum_{E} P(E) E \rho^0_{RQ} E^{\dagger} = \sum_{s,l} P(s)P(l|s) |s, B_l\rangle \langle s, B_l\ |\nonumber .
\end{eqnarray}
The states $|s, B_l\rangle$ are the common eigenstate of the stabilizers of the QEC code and the products $O^{X,Z}_{R_j} O^{X,Z}_{L_j}$, each of them denoting a syndrome and logical operator respectively.
For example, when $k=1$ there are four Bell states corresponding to the logical operators $I$, $O^X_L$, $O^Z_L$, and $O^X_L O^Z_L$, such that the label $l$ is in one-to-one correspondence with the logical Pauli operator.
The product $P(s)P(l|s) = P(s,l) \equiv \sum_{E\in \mathcal{E}_{s,l}} P(E)$, with $\mathcal{E}_{s,l}$ denoting the set of errors that generate the same syndrome $\vec{s}$ and logical quantum number $l$. $P(s)>0$ is the SPD and $P(l|s)$ the conditional probability of each logical operator $l$ given a syndrome $s$.   
Since $\rho_{RQ}$ is already diagonal we can rewrite the RCI as:
\begin{eqnarray}\label{eq:renyi_co_info}
    I^{(\alpha)} =  \frac{1}{1-\alpha}\log \left[\frac{\sum_s P^\alpha(s)\left(\sum_l P(l|s)\right)^\alpha}{\sum_{s} P^\alpha(s)\sum_l P^\alpha(l|s)}\right].
\end{eqnarray}
Here $P^{\alpha}(x)\equiv [P(x)]^{\alpha}$.
The argument of the logarithm contains the ratio of two quantities, $(\sum_l P(l|s))^\alpha = 1$ and $\sum_l P^\alpha(l|s)$, averaged over the unnormalized distribution $P^\alpha(s)$. This ratio quantifies, on average, how concentrated the conditional distribution $P(l|s)$ is over logical outcomes. As a result, Eq.~\eqref{eq:renyi_co_info} exhibits a singular behavior when $P(l|s)$ transitions from being sharply peaked at a single logical outcome to being distributed over multiple logical operators, as expected when crossing the QEC threshold \cite{terhal_quantum_2015,iyer_hardness_2015}. As observed in Ref.~\cite{fan_diagnostics_2024}, the RCI displays increasing thresholds with increasing $\alpha$, suggesting that logical fidelities sampled over the modified SPD
$Q_\alpha(s)= P^\alpha(s)/\sum_s P^\alpha(s)$
may show the same threshold behavior as the RCI.

\begin{figure}
    \centering
    \includegraphics[width=0.9\linewidth]{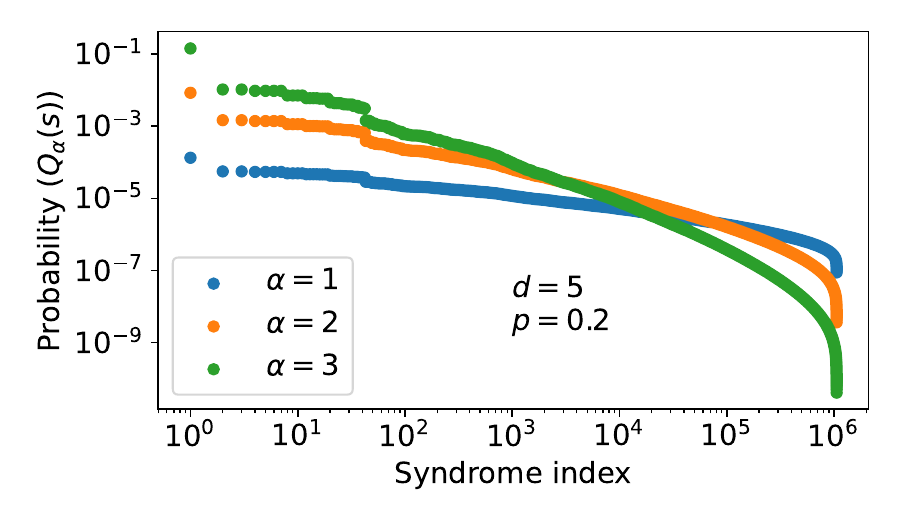}
    \caption{Numerical reconstruction of the modified syndrome distribution $Q_\alpha(s)$ for the $d=5$ unrotated surface code, $p=0.2$ and $\alpha=1,2,3$. We consider 20 stabilizers, which makes a total of $2^{20}$ possible syndromes. The index one denotes the trivial syndrome. The distribution becomes increasingly peaked around the more probable syndromes as $\alpha$ is increased.}
    \label{fig:synd_distribution}
\end{figure}

\begin{figure}[!h]
    \centering
    \includegraphics[width=\linewidth]{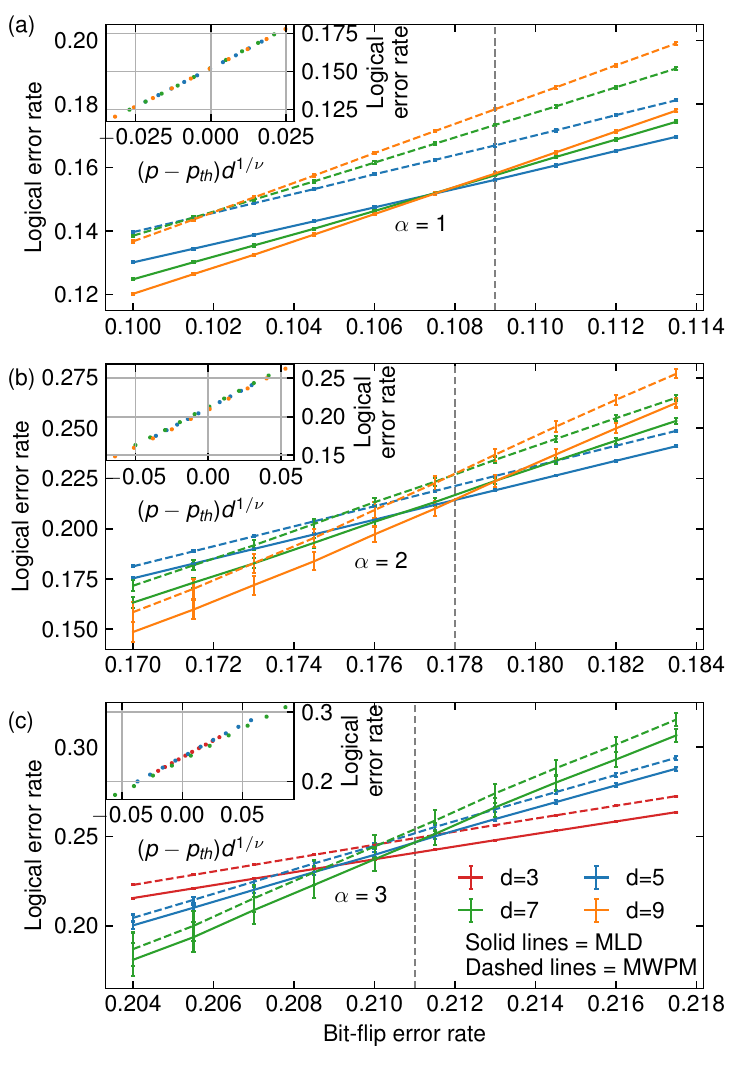}
    \caption{The logical error rate as a function of the bit-flip error rate in the vicinity of the respective threshold for (a) $\alpha = 1$, (b) $\alpha = 2$, and (c) $\alpha = 3$. The logical error rates are shown for both MWPM and MLD decoding and for multiple code distances. The insets show the collapsed data for MLD, with the extracted threshold $p_{th}$ and critical exponent $\nu$ values listed in Tab.~\ref{tab:comparison}. The dashed gray lines mark the corresponding RCI threshold values listed in Tab.~\ref{tab:comparison}.}
    \label{fig:thresholds}
\end{figure}

\emph{Rényi thresholds and powers of the syndrome probability distribution}: 
 For the toric code under bit-flip noise, the RCI displays a family of Rényi thresholds $p^{(\alpha)}_{\text{th}}$ \cite{fan_diagnostics_2024} for $\alpha=1,2,3,4,...,\infty$. Here $\alpha=1$ denotes the MLD threshold $p^{(1)}_{\text{th}}\sim0.109$ given by the phase transition of the random bond Ising model along the Nishimori line. For $\alpha>1$ the RCI is described by different spin models \cite{fan_diagnostics_2024}, all the way to $\alpha\rightarrow\infty$ representing the PS threshold \cite{english_thresholds_2025,english_ising_2025,smith_mitigating_2024}. Hence, obtaining the PS threshold $\alpha\rightarrow\infty$ in a QEC experiment implies to discard all non-trivial syndromes. As suggested by Eq.~\eqref{eq:renyi_co_info}, observing the thresholds for $1<\alpha<\infty$ then requires a \emph{resampling} of the data that effectively reproduces the modified SPD
$Q_\alpha(s) = P^\alpha(s)/\sum_s P^\alpha(s)$. 
To do so, we simulate the unrotated surface code \cite{dennis_topological_2002} under independent bit-flip noise $\mathcal{N}_i(\rho) = (1-p)\rho + p X_i \rho X_i$, with $i=1,...,n$. 
In Fig.~\ref{fig:synd_distribution} we show the syndrome distribution $Q_\alpha(s)$ for $\alpha = 1,2,3$ for the $d=5$ at a $20\%$ physical error rate. As $\alpha$ increases, the distribution becomes increasingly peaked around the most probable (trivial) syndrome $s_0$. This behavior is consistent with an increased threshold, as syndromes arising from correctable, low-weight errors become more probable \cite{chen_scalable_2025,sommers_spectral_2026}.

To compute the logical error rate $p_L^{(\alpha)}$ associated to $Q_\alpha(s)$, we first sample syndromes from $P(s)$. Then, for each measured syndrome $s$, we define a binary random variable $X_s$, where $X_s = 1$ if a logical error occurs after error correction and $X_s = 0$ otherwise. Hence $X_s$ depends on the decoding strategy. The logical error rate is computed as:
\begin{align} \label{eq:logical_error_rate}
    p^{(\alpha)}_L = \dfrac{\sum_{i=1}^{N}P^{\alpha-1}(s_i)X_{s_i}}{\sum_{i=1}^{N}P^{\alpha-1}(s_i)},
\end{align}
where $N$ is the number of samples. For $N\rightarrow \infty$ Eq.~\eqref{eq:logical_error_rate} yields the logical error rate sampled from $Q_\alpha(s)$. 
This method requires explicit evaluation of the probabilities $P(s)$. In general, this cannot be done efficiently. Below we discuss how $P(s)$ can be approximated from syndrome data. However, for the unrotated surface code under bit-flip noise, $P(s)$ can be computed using the efficient MLD decoder proposed in Ref.~\cite{bravyi_efficient_2014}, see Supplemental Material (SM) for details.
In Fig.~\ref{fig:thresholds} we show logical error rates $p_L^{(\alpha)}$ for $\alpha=1,2,3$ using SR. 
We compare two decoders: minimum weight perfect matching (MWPM) and the efficient MLD decoder proposed in Ref.~\cite{bravyi_efficient_2014}.
We extract the thresholds and the critical exponent $\nu$ from the numerical data using a finite-size scaling collapse, see Tab.~\ref{tab:comparison}.
Both decoders show increasing threshold with $\alpha$ and, as expected, MLD shows lower logical error rates than MWPM for every $\alpha$. 
The values obtained for MLD and those exhibited by the phase transition in RCI \cite{fan_diagnostics_2024} have relative differences below $2\%$, indicating that the resampled data indeed captures the same threshold and critical behaviour. 
Importantly, the transition in the RCI Eq.~\eqref{eq:def_coherent_information} is mapped to transitions in the 2D random-bond Ising, Ising and Ashkin--Teller models for $\alpha=1,2,3$ respectively, for which the critical exponents are known \cite{Delfino_2025,Cardy_1980}.  
For $\alpha=3$ we observe a mismatch between the exact and numerical critical exponents. This discrepancy is likely due to the limited system sizes and non-negligible logarithmic corrections in the Ashkin--Teller model \cite{Cardy_1980,Salas_1997}. 
Besides, the standard deviation of the estimator Eq.~\eqref{eq:logical_error_rate} for large $\alpha$ remains large even for large sample sizes, see SM for more details.
In the SM, we also show numerical evidence of a similar threshold enhancement in a quantum low-density parity check code (qLDPC) code, the lift-connected surface code \cite{old_lift-connected_2024}.

\begin{table}
    \centering
    \begin{tabular}{|c|c|ccc|}
        \hline
        & & $\alpha=1$ & $\alpha=2$ & $\alpha=3$ \\
        \hline
        \multirow{2}{*}{Threshold ($\%$)} & Numerical & $10.75(2)$ & $17.74(3)$ & $20.93(2)$ \\
        & RCI & $10.93(2)$ & $\sim17.8$ & $\sim21.1$ \\
        \hline
        \multirow{2}{*}{Exponent $\nu$} & Numerical & $1.53(6)$ & $0.99(6)$ & $0.83(6)$ \\
        & Exact RCI & $3/2$ & $1$ & $2/3$ \\
        \hline
    \end{tabular}
    \caption{The numerical threshold values and critical exponents for $\alpha=1,2,3$ obtained from the finite-size scaling collapse shown in Fig.~\ref{fig:thresholds}. The RCI values are determined from the statistical mechanical mapping of the RCI \cite{Merz_2002, fan_diagnostics_2024,Cardy_1980,Delfino_2025}. The values for MWPM are reported in the SM.}
    \label{tab:comparison}
\end{table}

\begin{figure}
    \centering
    \includegraphics[width=\linewidth]{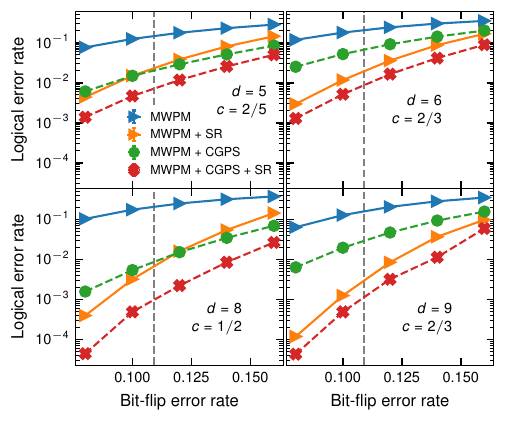}
    \caption{Logical error rate for surface code for bit-flip noise error rate $p=0.08,0.1,0.12,0.14,0.16$. We use MWPM only (blue), syndrome resampling (SR) (orange), complementary gap post-selection (CGPS)(green) and both SR and CGPS  (red). For SR we set $\alpha=2$. The complementary gap confidence parameter is set to $c=2/5,2/3,1/2,2/3$ \cite{cg_parameters} for code distances $d=5,6,8,9$ respectively. The lowest logical error rate is always obtained by decoding using both strategies. The grey dashed line denotes the standard optimal threshold $p\sim0.109$.}
    \label{fig:resampling_and_cg}
\end{figure}

\begin{figure}
    \centering
    \includegraphics[width=\linewidth]{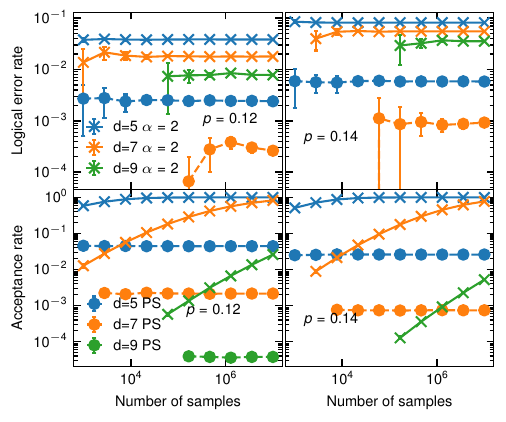}
    \caption{Logical error rate (top panels) and acceptance rate (bottom panels) of the resampling method under increasing number of samples $N$ for $p=0.12$ (left panels) and $p=0.14$ (right panels). Post-selected (circles) and $\alpha=2$ (stars) resampled data are considered. The acceptance rate is calculated as the number of samples kept in independently sampled batches of size $N$. The post-selected error rate for $d=9$ (not shown) is too low to be reliably computed with the present number of samples. Error bars are computed via bootstrapping and represent the $67\%$ confidence interval.}
    \label{fig:resampling_acceptance_rate}
\end{figure}

\emph{Syndrome resampling and decoding-based post-selection}: 
We now consider the typical situation in which $P(s)$ cannot be computed exactly. 
We consider a QEC experiment producing $N$ samples of syndromes $\vec{\sigma}=\{\sigma_1,\sigma_2,\ldots,\sigma_N\}$, where each syndrome $\vec{s}$ occurs with probability $P(s)$. The task is to reconstruct $P^\alpha(s)$ for $\alpha>1$ from the sampled data $\vec{\sigma}$.
This can be achieved by constructing the empirical distribution $\widehat{P}^{\alpha} =\binom{c_i}{\alpha}/\binom{N}{\alpha}$, where $c_i$ denotes the number of occurrences of syndrome $s_i$ in $\vec{\sigma}$. In the limit $N\to\infty$, $\widehat{P}^\alpha(s)$ converges to the exact distribution $P^\alpha(s)$. For finite $N$, a syndrome must appear at least $\alpha$ times to contribute to $\widehat{Q}_\alpha(s) = \widehat{P}^\alpha(s)/\sum \widehat{P}^\alpha(s)$. As a result, syndrome resampling is particularly effective when the SPD is strongly peaked, as in the low-error regime. In contrast, when $P(s)$ approaches a flat distribution, corresponding to high error rates, the required sample size scales exponentially with the number of qubits $n$ (see SM).
In Fig.~\ref{fig:resampling_and_cg} we see how resampling reduces the logical error rate of the rotated surface code \cite{tomita_low-distance_2014} under bit-flip noise, even for physical error rates exceeding the thresholds of MWPM and MLD \cite{dennis_topological_2002}. In this procedure, syndromes appearing fewer than $\alpha$ times are discarded, and decoding is performed on the remaining samples according the the distribution $\widehat{Q}_{\alpha}(s)$. As more data are collected, fewer syndromes are discarded, and in the limit $N\to\infty$ the discard rate vanishes. 
In Fig.~\ref{fig:resampling_acceptance_rate} we show how the logical error rate of MWPM and acceptance rate, i.e.~the number of samples kept within a set $\sigma$ of samples, are affected by considering only a finite number $N$ of samples for error rates above the MWPM threshold \cite{dennis_topological_2002}. First, the minimum $N$ for at least one syndrome to appear $\alpha$ times is exponentially large in the number of syndromes (see SM for details), but once this minimum value is attained, the acceptance rate is steadily increasing towards unity. We observe that the logical error rate converges faster to its steady value than the acceptance rate. 
Second, we compare SR $\alpha=2$ to PS. The number of samples needed for PS is higher, as consistent with being the limit $\alpha\rightarrow\infty$. This means  that there is a window of error rate and code distance where PS is unfeasible while SR is implementable and effective. 

We further consider complementary gap post-selection (CGPS) \cite{smith_mitigating_2024}. The complementary gap $\Delta$ is defined as the difference between the weight of the most likely error chain identified by MWPM and that of the most likely chain belonging to a different logical class. A sample is discarded if $(1-\Delta/d)>c$, where $c$ is a confidence parameter that is to be chosen. The limits $c=0$ and $c=1$ correspond to discarding all nontrivial syndromes and discarding none, respectively, analogous to the limits $\alpha\to\infty$ and $\alpha=1$ in our method. Unlike SR, each syndrome is deterministically either kept or discarded every time it appears, resulting in a finite discard rate as $N\to\infty$. As shown in Fig.~\ref{fig:resampling_and_cg}, CGPS also reduces the logical error rate in the regime considered.
We finally combine both strategies. We first construct $\widehat{Q}_\alpha(s)$, discard syndromes appearing fewer than $\alpha$ times, and resample the remaining data. On the selected samples, we apply MWPM decoding followed by CGPS with confidence parameter $c$. The resulting logical error rate is consistently lower than that obtained using either method alone and can be up to four orders of magnitude smaller when compared to bare MWPM decoding (see blue and red curves in Fig.~\ref{fig:resampling_and_cg}). The two mechanisms suppress logical errors through complementary effects: CGPS exploits decoder confidence, while SR enhances the weight of high-probability syndromes. Our results indicate that, at least for the surface code, the two methods act on substantially different subsets of syndromes.

\begin{figure}
    \centering
    \includegraphics[width=0.9\linewidth]{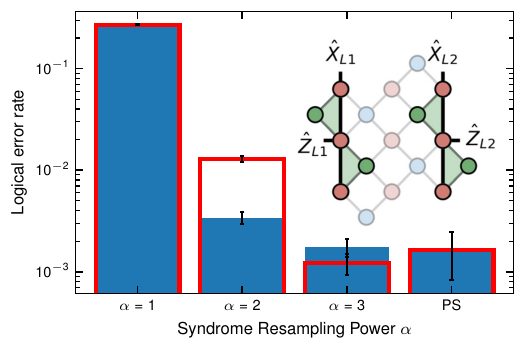}
    \caption{Logical error rate of the $\hat{Z}_{L_1}\hat{Z}_{L_2}$ expectation value in the repetition-code lattice-surgery experiment of Ref.~\cite{besedin_lattice_2026}. A surface-17 superconducting processor is initialized in a surface-code state, and the central column of data qubits is measured out, yielding two entangled three-qubit repetition codes encoding a logical Bell pair. Here, $\alpha=1,2,3$ denotes sampling from $\widehat{Q}_\alpha$ ($\alpha=1$ corresponds to no resampling), followed by MWPM decoding. PS denotes post-selection on the trivial syndrome. The data set contains $44,268$ samples. Red bars show SR using an SPD approximated from simulations rather than experimental data.}
    \label{fig:splitting_experiment}
\end{figure}

\emph{Application to experimental data}:
To demonstrate applicability in realistic settings, we benchmark SR on the recent three-qubit repetition-code lattice-surgery experiment of Ref.~\cite{besedin_lattice_2026}, shown in Fig.~\ref{fig:splitting_experiment}. Here, syndromes are replaced by detection events in the decoding graph, and the SPD is reconstructed directly from experimental data. For $\alpha=2$, SR reduces the logical error rate by approximately two orders of magnitude while retaining $40\%$ of the samples, compared with only $5\%$ under PS, thus achieving comparable performance at substantially lower sampling overhead. This improvement persists in the presence of experimental error mechanisms such as correlated errors, leakage, and crosstalk, demonstrating that SR is largely agnostic to microscopic noise details. We also apply SR using an SPD approximated from simulations (see SM), obtaining more than an order-of-magnitude reduction in logical error rate for $\alpha=2$ and near-optimal performance (within error bars) for $\alpha=3$. Thus, SR remains effective even when the SPD is estimated from a simplified noise model rather than directly from experimental data.

\emph{Discussion}:
We propose SR as a general method to enhance decoder thresholds by resampling syndrome probabilities. By down-weighting low-probability syndromes, SR suppresses high-weight errors and reduces logical error rates even above the standard ($\alpha=1$) error-correction threshold. We show that the resulting Rényi thresholds correspond to phase transitions in the RCI and introduce a method to approximate the SPD directly from syndrome data.
Unlike complementary-gap methods~\cite{english_ising_2025,english_thresholds_2025,smith_mitigating_2024}, SR requires no decoder-dependent decisions and applies to arbitrary stabilizer codes. Combined with CGPS, it further reduces logical error rates. Applied to experimental QEC data, SR achieves performance comparable to PS while retaining an order of magnitude more samples. Because SR relies only on syndrome probabilities, it naturally incorporates all error mechanisms reflected in the syndrome statistics, without requiring knowledge of the underlying error processes.
Our results suggest that existing QEC experiments across platforms can benefit from SR without additional decoding complexity or code-specific optimization: once the SPD is available, SR can enhance logical error rates. Although exact reconstruction of the SPD scales exponentially with system size, our experimental results show that accurate reconstruction is not required for effective logical-error suppression. Approximate SPDs can instead be obtained from inexpensive simulations or existing characterization methods~\cite{spitz_adaptive_2018,takou_estimating_2025,chen_calibrated_2022,sivak_optimization_2024,sundaresan_demonstrating_2023,remm_experimentally_2026}, mitigating this exponential cost.
We therefore envision SR as a simple and broadly applicable route to improving logical fidelities in near-term QEC experiments and algorithms.

\emph{Acknowledgment}: 
We thank Sam Smith, Ben Brown and Stephen Bartlett for providing the code implementing the complementary gap calculation. 
We thank Andreas Wallraff's Quantum Device Lab team at ETH Zurich for  providing the experimental raw data underlying the lattice surgery experiment of Ref. \cite{besedin_lattice_2026}.
We gratefully acknowledge funding by the U.S. ARO Grant No. W911NF-21-1-0007,  funding from the European Union’s Horizon Europe research and innovation programme under grant agreement No 101114305 (“MILLENION-SGA1” EU Project), and the German Federal Ministry of Research, Technology and Space (BMFTR) as part of the Research Program Quantum Systems, research project 13N17317 (”SQale”), and MUNIQC-ATOMS (Grant No. 13N16070), and by the Munich Quantum Valley (K-8), which is supported by the Bavarian state government with funds
from the Hightech Agenda Bayern Plus.  
M.M. furthermore acknowledges funding from
the ERC Starting Grant QNets through Grant No. 804247.
We also acknowledge support for the research that was sponsored by IARPA and the Army Research Office, under the Entangled Logical Qubits program, and was accomplished under Cooperative Agreement Number W911NF-23-2-0216. The views and conclusions contained in this document are those of the authors and should not be interpreted as representing the official policies, either expressed or implied, of IARPA, the Army Research Office, or the U.S. Government. The U.S. Government is authorized to reproduce and distribute reprints for Government purposes notwithstanding any copyright notation herein. 
We furthermore acknowledge support from the Deutsche Forschungsgemeinschaft (DFG, German Research Foundation) under  Germany’s Excellence Strategy Cluster of Excellence Matter and Light for  Quantum Computing (ML4Q) EXC 2004/1 390534769. The
authors gratefully acknowledge the computing time provided to them at the NHR Center NHR4CES at RWTH
Aachen University (Project No. p0020074). This is
funded by the Federal Ministry of Education and Research and the state governments participating on the
basis of the resolutions of the GWK for national high
performance computing at universities.

\emph{Data availability}: 
Data and codes for producing Fig.~\ref{fig:splitting_experiment} and for doing syndrome resampling on surface code stabilizer simulations are available in \url{https://doi.org/10.5281/zenodo.19232656} .

\bibliography{references_old,footnote}

\clearpage

\appendix

\section{SUPPLEMENTAL MATERIAL}

\section{Logical error rate for exact computation of SPD}\label{sec:exact_comp}

To compute the logical error rate $p_L^{(\alpha)}$ for the unrotated surface code under bit-flip noise, we use the estimator defined in Eq.~\eqref{eq:logical_error_rate}. The evaluation of this estimator requires the exact calculation of the syndrome probabilities $P(s)$, which we perform using the algorithm of Bravyi et al.~\cite{bravyi_efficient_2014}. The first step of this algorithm is to express the syndrome probability as a sum of two partition functions,
\begin{align}
    P(s) = \mathcal{Z}_I(s) + \mathcal{Z}_X(s),
\end{align}
where $\mathcal{Z}_I(s)$ is proportional to the partition function of the random-bond Ising model for a fixed bond configuration labelled by the syndrome $s$ \cite{dennis_topological_2002}. In $\mathcal{Z}_X(s)$ additional bonds are flipped along a logical X operator. These partition functions can be computed efficiently by mapping the two-dimensional classical model to a discrete, non-unitary time evolution of a one-dimensional quantum spin chain. Since this time evolution contains only matchgates \cite{Bravyi_2005,Terhal_2002}, it can be efficiently simulated on a classical computer. We refer the reader to Ref. \cite{bravyi_efficient_2014} for further details.

The variance of the estimator in Eq.~\eqref{eq:logical_error_rate} can be expressed as:
\begin{align} \label{eq:importance_sampling_error}
    (\Delta p^{(\alpha)}_L)^2 = \Big(\sum_{i=1}^N\dfrac{X_{s_i}}{N}\Big)\Big(1-\sum_{i=1}^N\dfrac{X_{s_i}}{N}\Big)\dfrac{\sum_{i=1}^{N}P^{2\alpha-2}(s_i)}{[\sum_{i=1}^{N}P^{\alpha-1}(s_i)]^2}.
\end{align}
For $\alpha=1$ this scales with $1/N$. For larger values of $\alpha$ the scaling depends on the shape of the distribution $P^{\alpha-1}(s)$, and the ideal $1/N$ behaviour is recovered only when $P^{\alpha-1}(s)$ is sufficiently flat. As discussed in the main text (see Fig.~\ref{fig:synd_distribution}), this condition is not satisfied in our case: for reasonable error rates ($p<0.5$) the syndrome distribution becomes increasingly peaked around $s_0$ as $\alpha$ increases. This also implies that the estimator in Eq.~\ref{eq:logical_error_rate} is more accurate for larger error rates as the syndrome distribution is less peaked in this regime. As a result of the large variance we are limited to simulate only small code distances (up to $d=9$ for $\alpha=2$ and $d=7$ for $\alpha=3$), with acceptable statistical uncertainty.

\section{MWPM thresholds}

\begin{figure}[!h]
    \centering
    \includegraphics[width=0.95\linewidth]{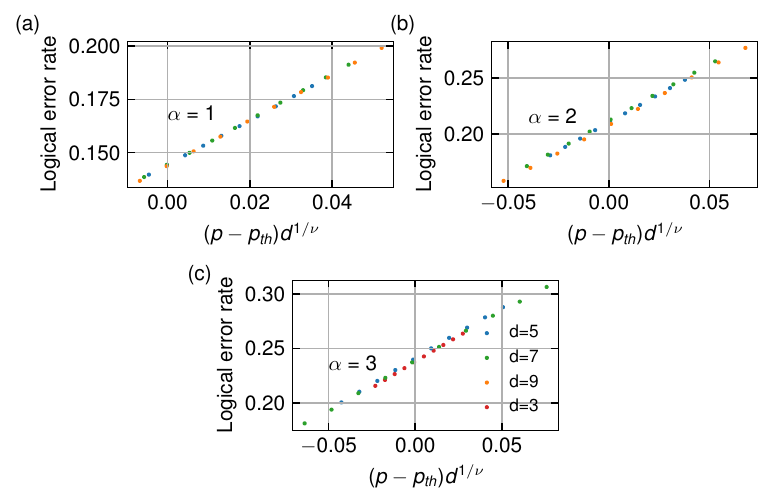}
    \caption{The logical error rate is shown as a function of the scaling variable $(p - p_{th}) d^{1/\nu}$ for MWPM decoding. The collapse is shown for $\alpha = 1, 2,$ and $3$ in panels (a), (b), and (c), respectively.}
    \label{fig:mwpm_collapse}
\end{figure}

\begin{table}[!h]
    \centering
    \begin{tabular}{|c|ccc|}
        \hline
        & $\alpha=1$ & $\alpha=2$ & $\alpha=3$ \\
        \hline
       Threshold ($\%$) & $10.15(2)$ & $17.58(4)$ & $21.01(3)$ \\
        \hline
        Exponent $\nu$ & $1.49(5)$ & $1.00(9)$ & $0.83(7)$ \\
        \hline
    \end{tabular}
    \caption{The numerical threshold values and critical exponents for $\alpha=1,2,3$ for MWPM decoding.}
    \label{tab:mwpm}
\end{table}

Here we present the finite-size scaling collapse of the MWPM logical error rates shown in Fig.~\ref{fig:thresholds}. We use the same method as for MLD decoding. The collapsed data for $\alpha = 1, 2, 3$ are shown in Fig.~\ref{fig:mwpm_collapse}, and the extracted threshold values and critical exponents are summarized in Tab.~\ref{tab:mwpm}. For $\alpha=1$ MWPM has clearly a smaller threshold than MLD. However for $\alpha=2,3$ both decoders seem to have similar thresholds within statistical uncertainty.  
The closing of the performance gap between both decoders can be explained by the fact that as $\alpha$ approaches the limit $\alpha\rightarrow\infty$, we enter the post-selection regime. In this limit all decoders have the same performance since one mostly sees the trivial syndrome. Therefore samples where MLD and MWPM output different corrections might be heavily suppressed already for $\alpha=2$. 

\section{Approximating powers of distributions from data}
\label{app:power_sampling}

In this section we describe in detail how to sample the power distribution $P^{\alpha}(x)/\sum_\alpha P^\alpha(x)$. We consider a generic random variable $x$ and make comments related to syndrome resampling in QEC.  
Let $X$ be a discrete random variable taking values in a finite or countable set
$\mathcal{X}$, with probability distribution $P(x)$.
For a fixed integer $\alpha \ge 2$, define the \emph{power distribution}
\begin{equation}
Q_\alpha(x) \;:=\; \frac{P^\alpha(x)}{Z_\alpha},
\qquad
Z_\alpha := \sum_{x \in \mathcal{X}} P^\alpha(x) .
\end{equation}
The quantity $Z_\alpha$ is the $\alpha$-th power sum of $P$ and determines the Rényi entropy
$H_\alpha(P) = \frac{1}{1-\alpha}\log Z_\alpha$.
In the following we describe how $Q_\alpha$ can be approximated and sampled from using
only an i.i.d.\ batch of samples from $P$.

Let $\vec{X} = \{ X_1,\dots,X_N \}\stackrel{iid}{\sim} P$, and denote by
\begin{equation}
c_x := \sum_{i=1}^N \mathbf{1}\{X_i = x\}
\end{equation}
the number of occurrences of symbol $x$ in the batch.
For integer $\alpha \ge 2$, define the empirical estimator
\begin{equation}\label{eq:approx_p}
\widehat{P}^{\,\alpha}(x)
\;:=\;
\frac{\binom{c_x}{\alpha}}{\binom{N}{\alpha}},
\qquad
\widehat{Z}_\alpha := \sum_{x \in \mathcal{X}} \widehat{P}^{\,\alpha}(x).
\end{equation}

A classical combinatorial identity due to Good~\cite{Good1953} yields
\begin{equation}
\mathbb{E}\!\left[\binom{c_x}{\alpha}\right] = \binom{N}{\alpha} P^\alpha(x),
\end{equation}
and therefore
\begin{equation}
\mathbb{E}\!\left[\widehat{P}^{\,\alpha}(x)\right] = P^\alpha(x),
\qquad
\mathbb{E}[\widehat{Z}_\alpha] = Z_\alpha .
\end{equation}
Hence $\widehat{P}^{\,\alpha}(x)$ and $\widehat{Z}_\alpha$ are \emph{unbiased} estimators of the
un-normalized power distribution and its normalization constant, respectively.
An empirical approximation to $Q_\alpha$ is obtained by normalization:
\begin{equation}
\widehat{Q}_\alpha(x) := \frac{\widehat{P}^{\,\alpha}(x)}{\widehat{Z}_\alpha}.
\end{equation}
While $\widehat{Q}_\alpha$ is biased due to the ratio form, it is consistent and converges
in total variation to $Q_\alpha$ as $N \to \infty$.

Regarding sampling error, both $\widehat{P}^{\,\alpha}(x)$ and $\widehat{Z}_\alpha$ are unbiased estimators of order $\alpha$.
Consequently, both quantities converge to their population values at rate
$O(N^{-1/2})$. 
Regarding the minimum number of samples needed to approximate $\widehat{Q}_{\alpha}(x)$, this amount is related to the probability that a value $x$ is observed $\alpha$ times.
A lower bound can be estimated by considering the element $x$ with the highest probability $P_{\text{max}} := \max{P(x)}$. The number of samples needed is then lower bounded as:
\begin{equation}
N \gtrsim \frac{\alpha}{P_{\text{max}}}.
\end{equation}
For our problem the trivial syndrome is always the one with the largest probability, such that $P_{\text{max}} \sim (1-p)^n$. Let us consider now two limits: low and high error rate regime. For $p\ll1$ and arbitrary $n$ we then obtain:
\begin{equation}
N \gtrsim \alpha e^{np}.
\end{equation}
The number of samples needed grows exponentially in $np$ which can turn to a linear dependence if $np\ll1$.
On the other hand, in the high error rate regime $p\sim0.5$ which can be a approximated as an uniform probability $P_{\text{max}}\sim 1/2^{n-k}$.
Thus the required number of samples scales as 
\begin{equation}
N \gtrsim \alpha 2^{n-k}.
\end{equation}
However, the syndrome probability distribution in relevant regimes is somewhere between those two regimes, therefore the number of samples needed to approximate $Q_{\alpha}(x)$ may vary a lot depending on the QEC code and physical error rates. 

In summary, the workflow for computing averages over $\widehat{Q}_{\alpha}(x)$ looks as follows: 
given the set $\vec{X} = \{ X_1,\dots,X_N \}\stackrel{iid}{\sim} P$, the resampling for $\alpha>1$ we do:

\begin{enumerate}
    \item Discard all $x \in \mathcal{X}$ that appear less than $\alpha$ times in the set. 
    \item Compute $\widehat{Q}_\alpha(x)$ using Eq. \eqref{eq:approx_p} on the remaining samples. 
    \item Obtain a sample $\widehat{X}$ by drawing from the distribution $\widehat{Q}_\alpha(x)$.
    \item Given $\widehat{X}=x_i$, select $x_i$ at random from the set $\vec{X}$. 
    \item Repeat 3 and 4 $\hat{N}$ times. 
    \item Compute the desired quantity by averaging over the set $\vec{\hat{X}}=\{\hat{X}_1,\dots, \hat{X}_{\hat{N}}\}$
\end{enumerate}

\section{Lattice surgery simulation}
\label{app:simulation}

\begin{table}[!h]
    \centering
    \begin{tabular}{|c|c|c|c|}
        \hline
        Pauli error & $CZ$ & Meas. & Single qubit gates \\
        \hline
        $p_x$ & 0.00067 & 0.00271 & 0.00016 \\
        \hline
        $p_y$ & 0.00067 &  0.00271 & 0.00016 \\
        \hline
        $p_z$  & 0.00512  & 0.02033  & 0.00123 \\
        \hline
    \end{tabular}
    \caption{Idling channel error rates during $CZ$ and measurements. All idling channels during single qubit gates are approximated by a single Pauli channel (right-most column).}
    \label{tab:error_model}
\end{table}

In the stabilizer simulation of the lattice-surgery experiment of Ref.~\cite{besedin_lattice_2026}, we approximate measurement, gate, initialization, and idling errors as Pauli noise. We use a measurement error rate $p_m=0.0106$. Single-qubit gates are followed by a depolarizing channel with error rate $p_1=0.009$, while two-qubit gates are followed by a uniform two-qubit depolarizing channel with error rate $p_2=0.0223$. Idling noise is modeled by Pauli twirling the amplitude-damping channel. The full procedure, including gate and coherence times, is given in Ref.~\cite{besedin_lattice_2026}; here, we only require the idling-channel probabilities during single-qubit gates, two-qubit gates, and measurements, listed in Table~\ref{tab:error_model}. The Stim circuit and simulation code are available in the Zenodo repository at \url{https://doi.org/10.5281/zenodo.19232656}.
The experimental data in Fig.~\ref{fig:splitting_experiment} are decoded using MWPM. The corresponding decoding graph is obtained from the detector error model generated using the noise model described above.

\section{Application to qLDPC codes}
\label{app:ldpc}

\begin{figure*}[h]
    \centering
    \includegraphics[width=0.9\linewidth]{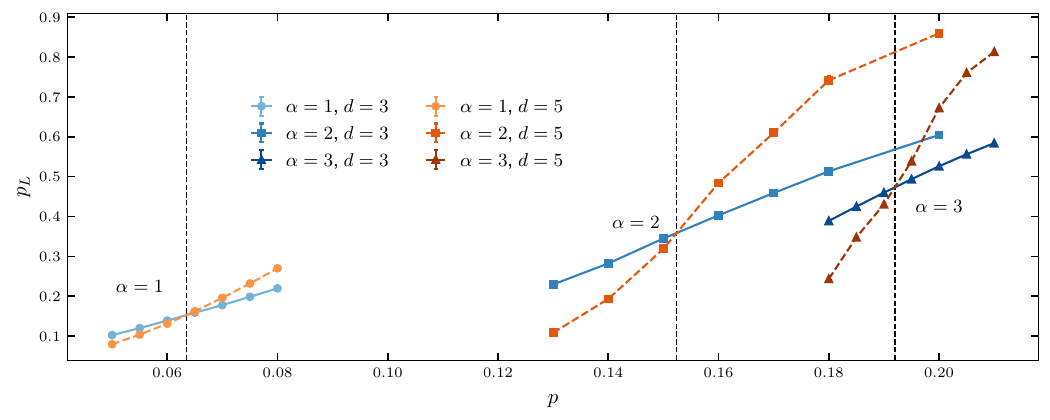}
    \caption{Logical error rate of the lift-connected surface (LCS) code using BP-OSD as a function of the bit-flip error rate $p$. The codes studied are $[[15,3,3]]$ and $[[65,5,5]]$. Blue and red curves correspond to $d=3$ and $d=5$, respectively. The color scales indicate syndrome resampling for $\alpha=1,2,3$, with $\alpha=1$ denoting no resampling. Crossings between $d=3$ and $d=5$ occur approximately at $p^{(1)}\sim 0.063$, $p^{(2)}\sim 0.152$, and $p^{(3)}\sim 0.191$, where the superscript denotes the corresponding value of $\alpha$.}
    \label{fig:lcs_code}
\end{figure*}

As a demonstration of the applicability of the SR method beyond the surface code, we consider a quantum low-density parity-check (qLDPC) code, the lift-connected surface (LCS) code~\cite{old_lift-connected_2024}. The LCS code is a lifted-product construction of repetition codes that yields $N$ unrotated surface codes of size $l+1$, joined by quasi 3D-local connections. The code parameters are $n=N[(l+1)^2+l^2]$, the code distance was shown to be $d=\min(N,2l+1)$, and the number of logical qubits is $k=N$. We consider two representatives with parameters $(N,l)=(3,1)$ and $(5,2)$, which yield the $[[15,3,3]]$ and $[[65,5,5]]$ codes, respectively. These are the smallest members of a family with the minimum possible number of physical qubits for increasing code distance. We consider bit-flip noise with probability $p$ and decoding using BP-OSD. For the decoder, we use the \texttt{ldpc} library with the same parameters as those given in Appendix B of Ref.~\cite{old_lift-connected_2024}. We approximate the distribution following the procedure described above in the subsection \textit{Approximating powers of distributions from data} and perform syndrome resampling. The logical error rate as a function of the physical error rate for different values of $\alpha$ is shown in Fig.~\ref{fig:lcs_code}. 
For $\alpha=1$, we observe a crossing between $d=3$ and $d=5$ at approximately $p^{(1)}\sim 0.063$, which is slightly lower than the value reported in Ref.~\cite{old_lift-connected_2024}. For $\alpha=2$ and $\alpha=3$, we observe crossings at $p^{(2)}\sim 0.152$ and $p^{(3)}\sim 0.191$, respectively. The latter results suggest a threshold enhancement similar to that observed for the surface code in the main text. This demonstrates that syndrome resampling effectively increases the threshold regardless of decoder performance and the specific structure of the QEC code.

\end{document}